\documentclass[a4paper,3p,times,twocolumn]{elsarticle}
\usepackage{latexsym,amsmath,amssymb,amsbsy}
\usepackage{graphicx}
\usepackage{nccfoots}
\usepackage[symbol]{footmisc}

\journal{Physics of Particles and Nuclei Letters, 2013}

\begin{document}


\title{Exposure of Nuclear Track Emulsion to $^8$He Nuclei\\
at the ACCULINNA Separator}

\author[1]{D.~A.~Artemenkov}
\author[1]{A.~A.~Bezbakh}
\author[1]{V.~Bradnova}
\author[1]{M.~S.~Golovkov}
\author[1]{A.~V.~Gorshkov}
\author[1]{P.~I.~Zarubin}
\author[1]{I.~G.~Zarubina}
\author[1,2]{G.~Kaminski}
\author[1]{N.~V.~Kondratieva} 
\author[1]{S.~A.~Krupko} 
\author[1]{K.~Z.~Mamatkulov}
\author[1]{R.~R.~Kattabekov}
\author[1]{V.~V.~Rusakova}
\author[1]{R.~S.~Slepnov}
\author[3]{R.~Stanoeva}
\author[1]{S.~V.~Stepantsov}
\author[1]{A.~S.~Vomichev}
\author[1,4]{V.~Chudoba}

\address[1]{Joint Institute for Nuclear Research, Dubna, Russia}
\address[2]{Institute of Nuclear Physics, Polish Academy of Sciences, Krakow, Poland}
\address[3]{South$\-$West University, Blagoevgrad, Bulgaria}
\address[4)]{Institute of Physics, Silesian University in Opava, Czech Republic}

\date{}

\begin{abstract} 
Nuclear track emulsion is exposed to a beam of radioactive $^8$He nuclei with an energy of 60 MeV and enrichment of about 80\% at the ACCULINNA separator. Measurements of 278 decays of the $^8$He nuclei stopped in the emulsion allow the potential of the $\alpha$  spectrometry to be estimated and the thermal drift of $^8$He atoms in matter to be observed for the first time.\par

\indent \par
\noindent \textbf{DOI:} 10.1134$/$S1547477113050026
\end{abstract}          

\maketitle

\begin{center}
INTRODUCTION
\end{center}

\indent At nuclear energies of a few MeV, it becomes possible to implant  radioactive  nuclei  into  the  detector material and thus investigate daughter states resulting from  their  decays  rather  than  the  implanted  nuclei themselves. For example, decays of light radioactive nuclei  can  populate  2$\alpha$ and 3$\alpha$-particle  states.  The known,  though  slightly  forgotten,  possibilities  of detecting  slow  nuclei  in  nuclear  track  emulsion  are worth considering in this connection. The advantages of this method are the best spatial resolution (about 0.5 $\mu$m), the possibility of observing tracks in the complete solid angle, and a record sensitivity range beginning with relativistic singly charged minimum ionizing particles. Nuclear track emulsion allows directions and ranges of beam nuclei and their decay products to be measured, which provides the basis for  $\alpha$ spectrometry.\par
\begin{figure*}[!ht]
\begin{center}
\tiny
\includegraphics[width=5in]{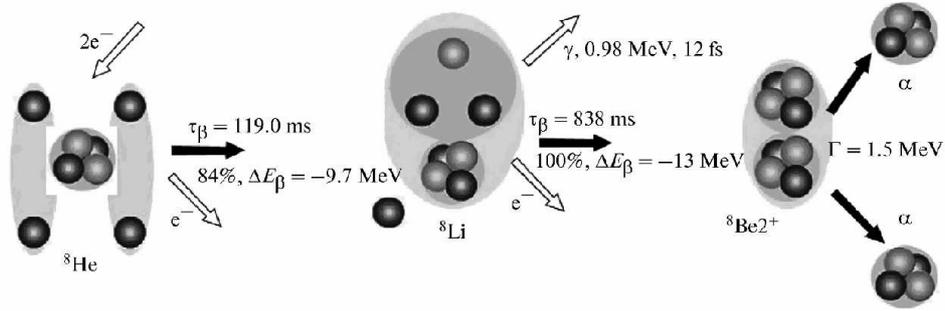}
\caption{Scheme of the main cascade decay channel for the $^8$He isotope. Circles are protons (light) and neutrons (dark). Darker background indicates clusters.}
\label{fig:1}
\end{center} 
\end{figure*} 
\indent More  than  50  years  ago,  hammerlike  tracks  of $^8$Be $\rightarrow$ 2$\alpha$ were  observed  in  nuclear  track  emulsion. They  resulted  from  $\beta$ decays  of  stopped  $^8$Li  and  $^8$B fragments produced in turn by high$\-$energy particles as emulsion  nuclei  underwent  splitting \cite{lib01}.  Another example is the first observation of the 2$\alpha$ $+$ \emph{p} decay of the $^9$C nucleus via the 2$^{+}$  state of the $^8$Be nucleus \cite{lib02}. Due  to  the  development  of  facilities  for  producing beams  of  radioactive  nuclei,  nuclear  track  emulsion turned out to be an effective tool for studying decays of light  exotic  nuclei  with  both  neutron  and  proton excess.\par
\indent As  a  first  step  within  this  approach,  the  nuclear track  emulsion  was  exposed  to  $^8$He  nuclei  with  an energy  of  $~$60  MeV  at  the  Flerov  Laboratory  of Nuclear Reactions (FLNR JINR) in March 2012. The features of  $^8$He decays are depicted in Fig.~1 in accordance with \cite{lib03}. After the  $^8$He nucleus is stopped and neutralized  in  the  substance,  the  formed  $^8$He  atom remains unbound (noble gas) and, as a result of thermalization, can drift in the substance until it undergoes  $\beta$  decay. The half$\-$life of the  $^8$He nucleus is  $\tau_{\beta}$ $=$ (119.0 $\pm$ 1.5) $\times$ 10$^{-3}$ s. This nucleus undergoes  $\beta$  decay to the 0.98$\-$MeV bound level of the  $^8$Li nucleus with a probability of 84\% and energy  $\Delta$\emph{E} $=$ 9.7 MeV. Then the $^8$Li nucleus with its half$\-$life  $\tau_{\beta}$ $=$ (838 $\pm$ 6) $\times$ 10$^{–3}$ s undergoes  $\beta$  decay to the 2$^{+}$  level of the  $^8$Be nucleus (3.03 MeV) with 100\% probability and energy  $\Delta$\emph{E} $=$ 13 MeV.  Finally,  the  $^8$Be  nucleus  decays  from  its  2$^{+}$ state with the width of 1.5 MeV to a pair of  $\alpha$  particles.\par
\indent Figure 2 shows a mosaic macrophotograph of the decay  of  the  $^8$He  nucleus  stopped  in  nuclear  track emulsion (one of several thousand events observed in this investigation). Video records of these decays made with a microscope and a camera are collected on the BECQUEREL site \cite{lib04}. This work deals with an analysis of the irradiation in question based on the measurements of 278 decays of this type.\par
\begin{figure}[!ht]
\includegraphics[width=0.45\textwidth]{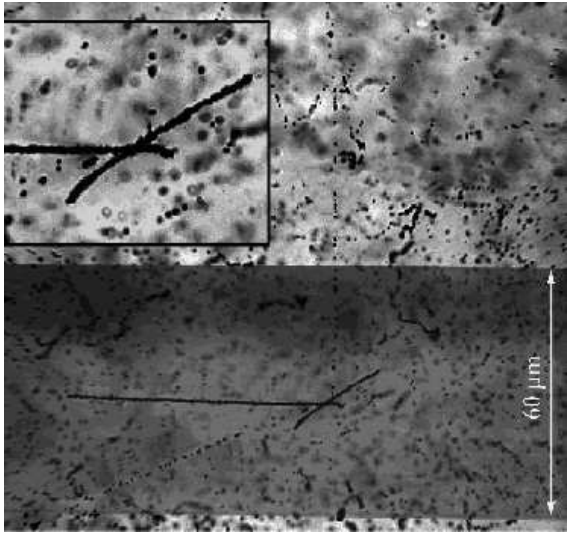}
\caption{Mosaic macrophotograph of the hammerlike decay of the $^8$He nucleus stopped in the nuclear track emulsion (horizontal track). The decay results in a pair of relativistic electrons (dotted tracks) and a pair of  $\alpha$  particles (oppositely directed short tracks). The inset shows the enlarged decay vertex. The decay image is superposed on the macrophotogtraph of a human hair 60 $\mu$m thick to illustrate the spatial resolution.}
\label{fig:2}
\end{figure} 

\begin{center}
EXPERIMENTAL
\end{center}

\indent Nuclear track emulsion was exposed to  $^8$He nuclei with an energy of 60 MeV at the ACCULINNA fragment separator of FLNR JINR \cite{lib05, lib06} (Fig.~3). A beam of heavy  $^{18}$O ions with an energy of 35 MeV$/$nucleon and intensity of $\sim $0.3\emph{p}$\mu$A extracted from the U400M cyclotron \cite{lib07} was used to produce  $^8$He nuclei. The  $^{18}$O ions bombarded  a  target  of  pyrolytic  graphite  175  mg$/$cm$^2$ thick  located  in  the  plane F$_{1}$.  The  target  was  a  disc 20 mm  in  diameter  and  1  mm  thick  sandwiched between two water$\-$cooled copper plates. The beam spot on the target was shaped with one of the plates used as a collimator 8 mm in diameter. This collimator was also used to tune the primary beam channel to the maximum  $^{18}$O beam transmission, usually as high as 90\%.\par
\begin{figure*}[!ht]
\begin{center}
\tiny
\includegraphics[width=5in]{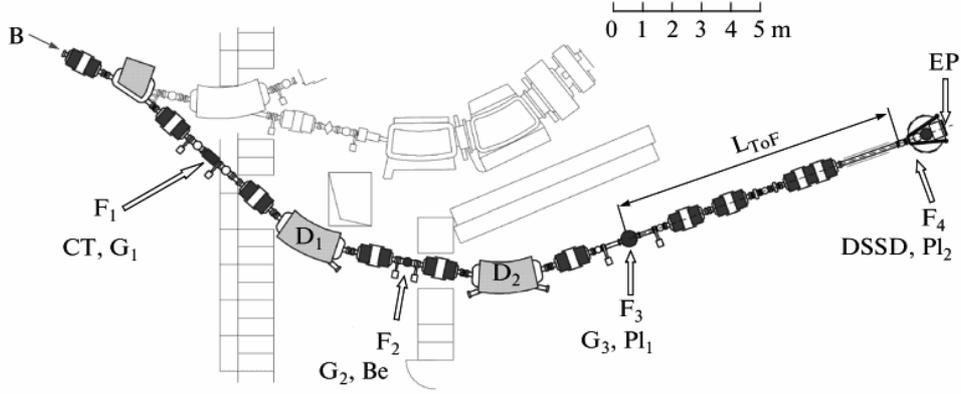}
\caption{Scheme showing the production of the 60-MeV  $^8$He beam at the ACCULINNA separator and the location of the nuclear track emulsion pellicles in the focus F$_{4}$  during their exposure to $^8$He nuclei. B is the direction of the primary beam extracted from the U400M accelerator; CT is the carbon target; F$_{1,2,3,4}$ are the focal planes; G$_{1,2,3}$  collimator gaps; Be is the beryllium wedge; Pl$_{1,2}$ are the plastic scintillator detectors; DSSD is the strip silicon detector; L$_{ToF}$ is the time-of-flight measurement path; and EP is the emulsion pellicle exposure place.}
\label{fig:3}
\end{center} 
\end{figure*} 
\indent The primary beam intensity was measured by two Faraday cups placed in the plane F$_{1}$ in front of and behind the collimator. The beam intensity was monitored during exposure by measuring the current on the tantalum foil (4~$\mu$m thick) fixed in place in front of the second Faraday cup. The parameters of the separator tuning for the production and shaping of the secondary  $^8$He  beam  in  the  achromatic  focus  F$_{3}$  and  final focus F$_{4}$ were determined from the field calculations for  the  dipole  and  quadrupole  elements  using  the TRANSPORT code \cite{lib08, lib09}. The beam composition in the final focal plane F$_{4}$ was set and monitored by (i) a gap with the dimensions \emph{X} $=$ $\pm$ 5~mm and \emph{Y} $=$ $\pm$ 10~mm and a beryllium wedge 1000  $\mu$m thick in the intermediate  plane  F$_{3}$, (ii)  a  gap  with  the  dimensions \emph{X} $=$ $\pm$ 5~mm and \emph{Y} $=$ $\pm$ 10~mm in the achromatic focus F$_{3}$, and (iii) two identical thin BC418 plastic scintillator detectors 60  $\times$ 40~mm in size and 127 $\mu$m thick viewed by two photomultiplier tubes on the left and on the right in F$_{3}$ and F$_{4}$  for the time-of-flight identification of particles and measurement of their energies. These detectors with a temporal resolution of about 0.5 ns (half-width at half-maximum) installed in the straight section 8.5~m long ensured particle energy determination with an accuracy no worse than 1\%. \par
\indent The design of the time-of-flight detector is shown in  Fig.~4. A  2-$\mu$m-thick  foil  of  aluminized  Mylar served as a reflector. Diffuse reflection is provided by the Tyvec light guide. The scintillators were viewed on the left and on the right downstream of the beam by two fast XP2020 photomultiplier tubes, which allowed correcting the signal amplitude and time dependences on  the  particle  entrance  point  in  the  detector.  The positional resolution of the detector in the horizontal coordinate determined from the relative left-to-right signal amplitude ratio was about 10~mm. The above dependences were especially noticeable and important for the detector in the focus F$_{4}$, where the converging secondary beam spot was an ellipse 40  $\times$ 30~mm in size.\par
\indent A  position-sensitive  silicon  detector  1~mm  thick with an active area 50 $\times$ 58~mm in size and 1.8-mm-wide sensitive strips was installed at a distance of 130~cm from the  scintillation  detector  in  F$_{4}$ downstream  of  the beam  immediately  in  front  of  the  vacuum  chamber exit  window.  The  silicon  detector  allowed  the  beam profile to be determined in two coordinates with an accuracy of 1.8~mm and particles to be uniquely identified by measuring the particle energy loss more accurately than with the plastic scintillator detector. Measured by this detector, the  $^8$He beam profile in the \emph{X} and  \emph{Y}  planes  was  about  26~mm  (half-width  at  half-maximum). Figure 5a presents an identification picture  of  the  beam  of  radioactive  nuclei  obtained  by measuring  the  energy  loss  of  particles  in  the  silicon detector as a function of their time of flight over the path  of  8.5~m  when  the  separator  was  tuned  to  the maximum  $^8$He beam transmission.\par
\begin{figure}[!ht]
\includegraphics[width=0.45\textwidth]{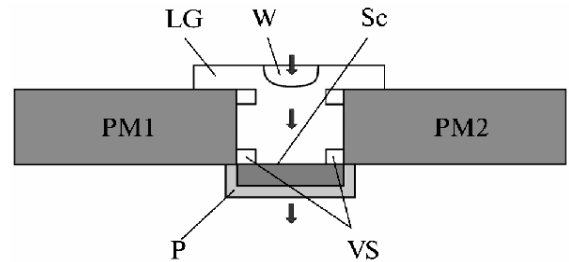}
\caption{Schematic  view  of  the  scintillation  detector  for measuring  the  time  of  flight  of  the  fragments  along  the straight section F$_{3}$–F$_{4}$ of ACCULINNA: W is the beam entrance window covered with a reflector; LG is the total diffuse  light  guide;  SC  is  the  scintillator;  PM$_{1,2}$  are  the photomultipliers; R is the reflector; VS are vacuum seals.}
\label{fig:4}
\end{figure} 
\indent The magnetic rigidity of the dipole magnets D$_{1}$ and D$_{2}$ \emph{B}$\rho_{1}/$\emph{B}$\rho_{2}$ $=$ 2.8903$/$2.829 T m and  the  wedgeshaped beryllium absorber (1~mm) with gaps of $\pm$ 5~mm in  maximum  dispersion  plane  F$_{2}$ set  the  following characteristics of the secondary  $^8$He beam in plane F$_{4}$: energy 23.8 $\pm$ 0.9 MeV$/$nucleon, intensity $\sim $50 particles$/$s at a primary beam intensity of $\sim$0.3\emph{p}$\mu$A, and $^8$He enrichment $\sim$80\% (Figs.~5b,~5c).\par
\indent Considering the detector material inside the vacuum  chamber,  the  Kapton  exit  window  (125~$\mu$m thick),  and  the  aluminum  plate  (3900~$\mu$m  thick) installed in the air behind the window at a distance of $\sim$2~cm, the calculated energy of the  $^8$He nuclei before their hitting the emulsion assembly was about 59.2 $\pm$ 4.5 MeV. Several emulsion pellicles were exposed to the beam with these characteristics. The exposure of each  pellicle  was  about  10 min  long,  which  corresponded  to  the  integral  flux  of  about  4 $\times$ 10$^4$ $^8$He nuclei.\par
\begin{figure*}[!ht]
\begin{center}
\tiny
\includegraphics[width=5in]{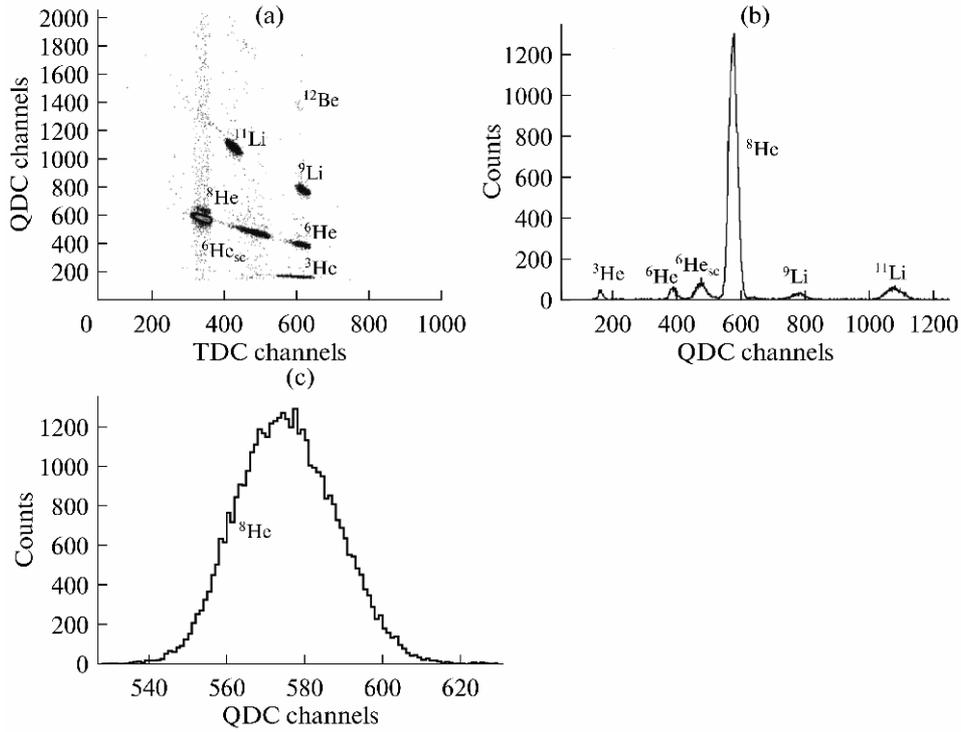}
\caption{Composition of the beam produced at the ACCULINNA separator tuned to the $^8$He isotope from the fragmentation of the $^{18}$O nuclei with an energy of 35 MeV$/$nucleon on the $^{12}$C target. (a) Identification of particles by the silicon detector and from the time of flight; (b) spectra of energy lost by all beam particles in the silicon detector 1 mm thick; and (c) energy loss of $^8$He nuclei alone. The sum of counts in (b) and (c) was used to find the beam enrichment in $^8$He nuclei.}
\label{fig:5}
\end{center} 
\end{figure*} 
\indent The  emulsion  used  for  exposure  (conventionally referred to as series 21) is an analogue of the known BR–2  emulsion  recently  reproduced  at  the  Mikron factory of the Slavich Company \cite{lib10}, which is sensitive to minimum ionizing relativistic particles. This investigation can be regarded as a calibration of the nuclear track emulsion in a physical experiment.
\indent To choose the optimum observation of  $^8$He stops, emulsion pellicles 9 $\times$ 12~cm in size and 107~$\mu$m thick produced by splashing onto the glass substrate 2~mm thick were placed in the beam both across it and at an angle to its axis (10 to 20$^{o}$). The subsequent scanning revealed that the best pellicle for analysis was the one inclined at an angle of 10$^{o}$. An inclination of the plate resulted in a larger deceleration layer in the emulsion pellicle. It is this pellicle that was used for analysis in this work. Before exposure, the pellicles were wrapped in two layers of black paper 100~$\mu$m thick each. The beam nuclei were thus given additional deceleration, especially sensitive at the angle of 10$^{o}$.\par

\begin{center}
ANALYSIS OF HAMMERLIKE DECAYS
\end{center}

\begin{figure}[!ht]
\includegraphics[width=0.45\textwidth]{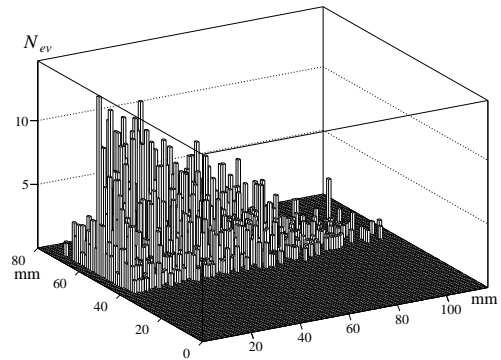}
\caption{Beam profile in the hammerlike decays; the bin size is 1 $\times$ 1 mm.}
\label{fig:6}
\end{figure} 
\indent As  the  pellicle  was  scanned  using  an  MBI–9 microscope with a 20~$\times$ lens, the primary search for $\beta$ decays  of  $^8$He  nuclei  was  focused  on  hammerlike events (Fig.~2). The absence of tracks of a decay electron in the observed event was  interpreted as a consequence of the inadequately effective observation of all decay tracks in the emulsion pellicle. The most problematic  background  for  selection  by  this  criterion could arise from decays of  $^8$Li nuclei. However, as follows from Fig.~5a, this isotope is not observed. Beta decay of stopped  $^9$Li nuclei with the formation of  $^8$Be and  the  emission  of  a  delayed  neutron  (probability $\sim$50\%)  could  also  meet  the  above  criterion,  but  the admixture of these nuclei is small (Fig.~5a). In addition, for the decay of the  $^8$Be 2$^{+}$ state to be hammerlike,  it  must  populate  the  $^9$Be  level  at  an  energy  no lower than 4.7 MeV. Otherwise, the decay proceeds via the 0$^{+}$  ground state of the  $^8$Be nucleus and is therefore hardly observable even in emulsion. Thus, the background  from  decays  of  $^8$Li  and  $^9$Li  nuclei  could  be ignored.\par
\indent There is often a gap observed between the stopping point  and  the  hammerlike  decay  itself.  These  "broken" events were attributed to the drift of thermalized $^8$He atoms that resulted from the neutralization of  $^8$He nuclei. This effect is determined by the nature of $^8$He, and  these  events  are  particularly  reliably  identified. Since  $^8$He nuclei dominate in the beam ($\sim$80\%), the distribution of the hammerlike decays over the emulsion  area  can  be  presented  jointly  for  all  observed events,  including  1413  "whole" and  1123  "broken" ones (Fig.~6). There is a uniform distribution of vertices in the vertical coordinate and a characteristic scatter, as a result of separation, in the horizontal coordinate.\par
\indent The events that included at least one electron were further  measured  using  the  90 $\times$ KSM  microscopes. The average length of the beam tracks for 136 whole events was  $<$ \emph{L} ($^8$He) $>$  $=$ 263 $\pm$ 11 $\mu$m at the root-mean-square  scatter  (RMS)  113~$\mu$m,  and  for  142  broken events it was 296 $\pm$ 10~$\mu$m at the RMS 118~$\mu$m. Since the difference in the parameters is insignificant, the distributions  of  ranges  in  these  events  are  jointly depicted in Fig.~7. The SRIM simulation program \cite{lib11} allows the kinetic energy of the  $^8$He nuclei that penetrated into the emulsion pellicle to be evaluated on the basis of the range measurements. Its average value is $<$\emph{E} ($^8$He)$>$ $=$ 29 $\pm$ 1 MeV at the RMS 10 MeV.\par
\indent The substantially lower average $^8$He energy and its larger spread at the entrance to the emulsion pellicle when compared with the value set by the fragment separator is due to the deceleration in the wrapping paper. The calculated average range in the emulsion  $<$\emph{L} ($^8$He)$>$ with  allowance  for  the  deceleration  in  1  mm  of  the paper is about 280~$\mu$m \cite{lib11}. In addition, the inhomogeneous structure of the paper contributes to the considerable spread of ranges \emph{L} ($^8$He) (Fig.~7), which calculations fail to describe. Thus, the inhomogeneity of the light-proof paper turns out to be a factor that cannot be ignored and, at the same time, is difficult to take into account exactly. The given estimate of the effective paper thickness can be a reference for planning irradiation with other nuclei.\par
\begin{figure}[!ht]
\includegraphics[width=0.45\textwidth]{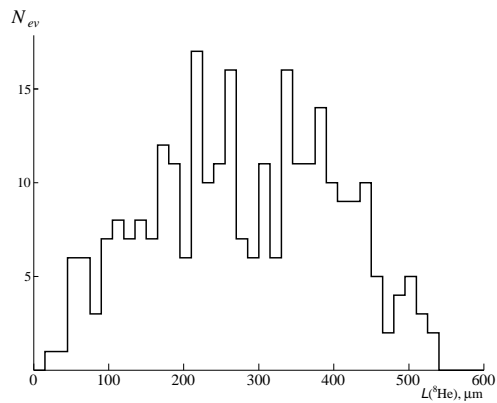}
\caption{Range distribution of the $^8$He tracks in the emulsion.}
\label{fig:7}
\end{figure} 
\begin{figure}[!ht]
\includegraphics[width=0.45\textwidth]{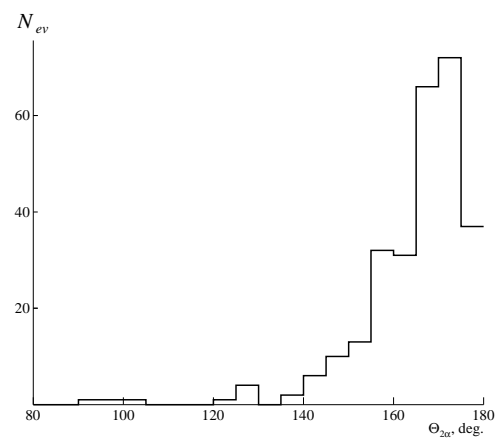}
\caption{Angle $\Theta_{2\alpha}$ distribution in pairs of $\alpha$ particles.}
\label{fig:8}
\end{figure} 
\begin{figure}[!ht]
\includegraphics[width=0.45\textwidth]{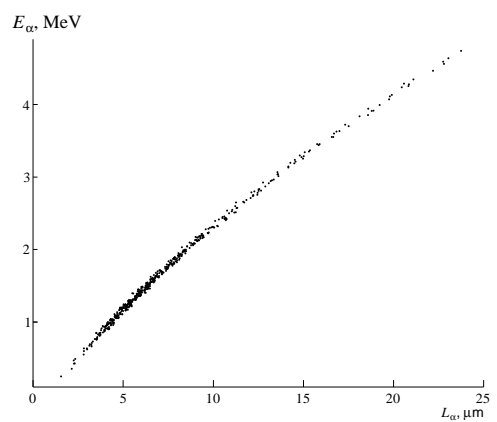}
\caption{Determination  of  the $\alpha$ particle  energy  from  the measured ranges.}
\label{fig:9}
\end{figure} 
\begin{figure}[!ht]
\includegraphics[width=0.45\textwidth]{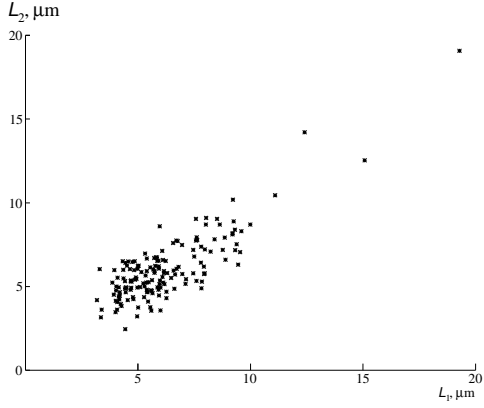}
\caption{Distribution of ranges \emph{L}$_1$ and \emph{L}$_2$ in pairs of $\alpha$ particles.}
\label{fig:10}
\end{figure} 
\begin{figure}[!ht]
\includegraphics[width=0.45\textwidth]{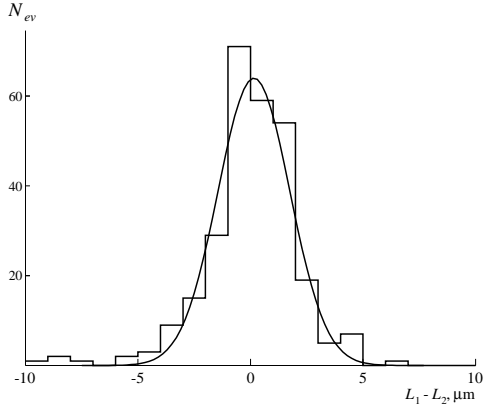}
\caption{Distribution of range differences \emph{L}$_1$ $\--$ \emph{L}$_2$ in pairs of $\alpha$ particles; the curve is the Gaussian function.}
\label{fig:11}
\end{figure} 
\indent Coordinates of decay vertices and stops of decay $\alpha$-particles were determined for the hammerlike decays from  136  whole  and  142  broken  events.  In  broken events the decay coordinate was found by extrapolating the electron track to the hammerlike track. Thus, the  emission  angles  and  ranges  of $\alpha$-particles  were obtained.\par
\begin{figure}[!ht]
\includegraphics[width=0.45\textwidth]{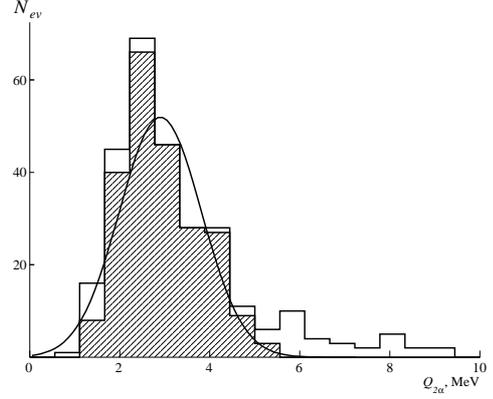}
\caption{Energy \emph{Q}$_{2\alpha}$ distribution  of $\alpha$ particle  pairs;  the hatched histogram satisfies the event selection conditions \emph{L}$_1$ and \emph{L}$_2$ $<$ 12.5~$\mu$m. The curve is the Gaussian function.}
\label{fig:12}
\end{figure} 
\begin{figure}[!ht]
\includegraphics[width=0.45\textwidth]{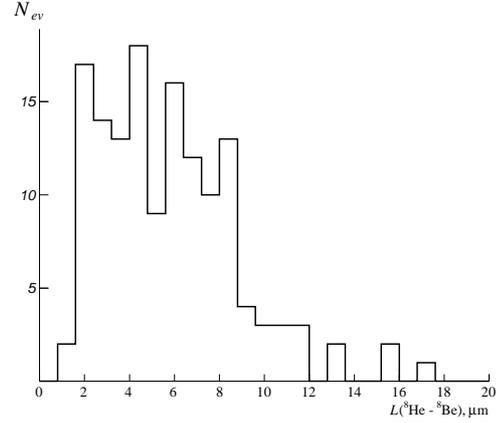}
\caption{Distribution of distances \emph{L}($^8$He $\-– ^8$Be) between the $^8$He stopping points and the $^8$Be(2$^+$) decay vertices in the broken events.}
\label{fig:13}
\end{figure} 
\indent Figure 8 shows the emission angle distribution of pairs of $\alpha$ particles. The average value of the angles is $<$ $\theta_{2_{\alpha}}$ $>$ $=$ (164.9 $\pm$ 0.7)$^{\circ}$ at the RMS (116 ± 0.5)$^{\circ}$. A small kink in  the  hammerlike  decays  is  determined  by  the momenta carried away by \emph{e}$\nu$ pairs. Figure 9 depicts the relation between the ranges \emph{L}$_{\alpha}$ of the $\alpha$-particles from the hammerlike decays and the energies \emph{E}$_{\alpha}$  found from the spline interpolation of the range–energy calculation within the SRIM model. The average of the  $\alpha$-particle ranges is 7.4 $\pm$ 0.2 $\mu$m at the RMS 3.8 $\pm$ 0.2 $\mu$m, which  corresponds  to  the  average  kinetic  energy $<$ \emph{E}($^4$He) $>$ $=$ 1.79 $\pm$ 0.03 MeV at the RMS 0.8 MeV. The ranges \emph{L}$_{1}$ and \emph{L}$_{2}$  of  $\alpha$  particles in pairs exhibit a distinct correlation (Fig. 10). The distribution of the range differences \emph{L}$_{1}$ $\-–$ \emph{L}$_{2}$ (Fig. 11) has the RMS 2.0 $\mu$m.\par
\indent Knowing the energy and emission angles of the $\alpha$ particles, we can obtain the $\alpha$ decay energy distribution \emph{Q}$_{2\alpha}$. The  relativistically  invariant  variable  \emph{Q} is defined as a difference between the invariant mass of the  final  system  \emph{M}$^*$ and  the  mass  of  the  primary nucleus \emph{M}; i.e., \emph{Q} $=$ \emph{M}$^*$ – \emph{M}. Here \emph{M}$^*$ is defined as a sum of all products of the fragment four-momenta \emph{P}$_{i,k}$, \emph{M}$^{*2}$ $=$ ($\Sigma$\emph{P}$_{j}$)$^2$ $=$ $\Sigma$(\emph{P}$_{i}$\emph{P}$_{k}$).\par
\indent The \emph{Q}$_{2\alpha}$ distribution (Fig.~12) mainly corresponds to the decays of $^8$Be nuclei from the excited 2$^{+}$ state. Its average value  $<$\emph{Q}$_{2\alpha}>$, however, turned out to be slightly greater than expected, which results from a small tail in  the  region  of  large  \emph{Q}$_{2\alpha}$ that  obviously  does  not fit into the description by the Gaussian function. Applying the selection criteria \emph{L}$_{1}$ and \emph{L}$_{2}$ $\prec$ 12.5~$\mu$m and  $\theta$ $\prec$ 145$^{\circ}$,  we  obtain  $<$\emph{Q}$_{2\alpha}$ $>$ $=$  2.9  $\pm$ 0.1  MeV  at  the  RMS 0.85 $\pm$ 0.07 MeV, which corresponds to the 2$^{+}$  state.\par
\indent The reason why the tail arises in the \emph{Q}$_{2\alpha}$ distribution is obscure and calls for further analysis. According to  Fig. 10,  the  ranges  \emph{L}$_{1}$ and \emph{L}$_{2}$ correlate  at  values greater than 12.5 $\mu$m as well. Therefore, an increase in ranges  cannot  be  attributed  to  fluctuations  of  ranges due to recombination of He$^{+2}$ ions. This fact should be taken into consideration in a comprehensive analysis.\par
\indent The resolution of the nuclear track emulsion is enough to find the distances \emph{L}~($^8$He $\--$ $^8$Be) between the $^8$He stopping points and the $^8$Be (2$^+$) decay vertices in the broken events  (Fig. 13). The average value $<$~\emph{L}~($^8$He $\--$ $^8$Be)~$>$ $=$ 5.8 $\pm$ 0.3 $\mu$m at the RMS 3.1 $\pm$ 0.2 $\mu$m can be associated with the average drift length of thermalized $^8$He atoms.\par
\indent The observation of the drift indicates the possibility of  generating  radioactive  $^8$He  atoms  and  pumping them out from sufficiently thin targets. The drift rate and  length  can  be  increased  by  heating  the  target. Extensive research in this direction with application to $^6$He  isotopes  is  under  way \cite{lib12, lib13}. The  prospect  of accumulating  considerable  amounts  of  $^8$He  atoms exists. Radioactive  $^8$He gas can be used for measuring the  $^8$He half - life at a new level of accuracy and for the laser  spectroscopy  of  $^8$He.  Of  applied  interest  is  the investigation of thin films by pumping  $^8$He atoms with their  particular  penetrating  power  and  depositing them onto detectors.\par

\begin{center}
CONCLUSIONS
\end{center}
\indent This  work  demonstrates  the  capabilities  of  the recently reproduced nuclear track emulsion exposed to a  beam  of  $^8$He  nuclei.  The  test  experiment  allowed radioactive  $^8$He nuclei to be independently identified by their decays as they stopped in the emulsion, the possibility of carrying out the  $\alpha$  spectrometry of these decays to be estimated, and the drift of thermalized  $^8$He atoms in matter to be observed for the first time. The experiment proved the high purity of the beam of radioactive nuclei  formed  at  the  ACCULINNA  facility  with  an energy ranging from 10 to 30 MeV$/$nucleon. The analysis of 278 decays of  $^8$He nuclei can be a prototype for investigating decays of  $^{8,9}$Li,  $^{8,12}$B,  $^9$C, and  $^{12}$N nuclei in which the  $^8$Be nucleus serves as a marker. The nuclear track emulsion can be used for the diagnostics of beams of radioactive isotopes.\par
\indent The statistics of the hammerlike decays observed in this work is a small fraction of the flux of  $^8$He nuclei, and the measured decays constitute 10\% of that fraction.  This  limitation  was  dictated  by  "reasonable" time  and  labor  expenditure.  At  the  same  time  the nuclear  track  emulsion  with  implanted  radioactive nuclei offers the basis for using automatic microscopes and  image-recognition  programs,  making  it  possible to   hope   for   unprecedented   statistics   of   analyzed decays. Thus, classical methodology can be synergistically combined with modern technologies.\par

\begin{center}
ACKNOWLEDGMENTS
\end{center}
\indent The  authors  are  grateful  to  the  project  curator O.I.~Orurk (Moscow);  Yu.A. Berezkina,  A.V. Kuznetsov,   and   L.B. Balabanova   (Micron,   Slavich, Pereslavl' Zalesskii)   for   making   new   samples   of nuclear   track   emulsion   available;   A.S. Mikhailov (Moscow  Cinematography  and  Video  Institute)  for methodological assistance in reproducing the nuclear track  emulsion  technology;  N.G. Polukhina (FIAN) and A.I. Malakhov (JINR) for constant assistance in our work, and the technician G.V. Stel'makh (JINR) for making an appreciable contribution to the search for events.\par
\indent The  work  was  supported  by  the  Russian  Foundation for Basic Research, project no. 12-02-00067, and by  grants  from  the  Plenipotentiaries  of  the  Governments of Bulgaria and Romania at JINR.\par


\indent \par
\indent \emph{Translated by M.~Potapov}

\begin{thebibliography}{99}

\bibitem{lib01} C.F.~Powell, P.H.~Fowler, and D.H.~Perkins, \emph{Study of Elementary Particles by the Photographic Method} (Pergamon, London, 1959), pp. 465–472.
\bibitem{lib02} M.S.Swami, J.Schneps, and W.F.Fry, Phys. Rev. \textbf{103}, 1134–1135 (1956).
\bibitem{lib03} F.~Ajzenberg-Selove, Nucl. Phys. A~\textbf{490}, 1-266 (1988); TUNL, Nuclear Data Evaluation Project. http://www.tunl.duke.edu/NuclData/
\bibitem{lib04} The BECQUEREL Project, http://becquerel.jinr.ru/miscellanea/8 He/8 He.html/
\bibitem{lib05} A.M.~Rodin et al., Nucl. Instrum. Methods Phys. Res., Sect. B~\textbf{204}, 114-118 (2003).
\bibitem{lib06} The ACCULINNA Project. http://aculina.jinr.ru/
\bibitem{lib07} Flerov Laboratory of Nuclear Reactions, U400M Accelerator Complex. http://flerovlab.jinr.ru/ flnr/ u400m.html
\bibitem{lib08} U.~Rohrer,  PSI  Graphic  Transport  Framework based on a CERN - SLAC - FERMILAB version by K.L.~Brown et al. http://aea.web.psi.ch/Urs Rohrer/MyWeb/trans.html/
\bibitem{lib09} K.L.~Brown et al., 1980 CERN Yellow Report 80-04.
\bibitem{lib10} Slavich Company. www.slavich.ru/
\bibitem{lib11} J.F.~Ziegler, J.P.~Biersack, and M.D.~Ziegler, \emph{SRIM - the  Stopping  and  Range  of  Ions  in  Matter} (SRIM  Co, 2008); Particle Interactions with Matter, SRIM - The Stopping and Range of Ions in Matter. http://srim.org/
\bibitem{lib12} A.~Knecht et al., Phys. Rev. C~\textbf{86}, 035506 (2012).
\bibitem{lib13} T.~Stora et al., Europhys. Lett.~\textbf{89}, 32001 (2012).
\end{thebibliography}
\end{document}